%% file: Template.tex
\documentclass{article}
\usepackage{spconf,amsmath,graphicx,hyperref}
\usepackage{float}
\usepackage{amsfonts}
\usepackage{multirow}
\ninept

\input{define}

\title{Towards Multimodal Query-Based Spatial Audio Source Extraction}
\name{Chenxin Yu\textsuperscript{1}, Hao Ma\textsuperscript{2}\sthanks{The work is done with TeleAI.}, Xu Li\textsuperscript{3}, Xiao-Lei Zhang\textsuperscript{2\dag}, Mingjie Shao\textsuperscript{4}, Chi Zhang\textsuperscript{2}, and Xuelong Li\textsuperscript{2\dag}\thanks{\textsuperscript{\dag}Corresponding authors: Xuelong Li (xuelong\_li@ieee.org), and Xiao-Lei Zhang (xiaolei.zhang@nwpu.edu.cn)}}

\address{%
  \small\textsuperscript{1}Department of Computer Science, Cornell University, USA\\
  \small\textsuperscript{2}Institute of Artificial Intelligence (TeleAI), China Telecom\\
  \small\textsuperscript{3}Kuaishou Technology, Beijing, China\\
  \small\textsuperscript{4}State Key Laboratory of Mathematical Sciences, AMSS, Chinese Academy of Sciences, Beijing, China
}

\begin{document}
\maketitle
\begin{abstract}
Query-based audio source extraction seeks to recover a target source from a mixture conditioned on a query. Existing approaches are largely confined to single-channel audio, leaving the spatial information in multi-channel recordings underexploited. We introduce a query-based spatial audio source extraction framework for recovering dry target signals from first-order ambisonics (FOA) mixtures. Our method accepts either an audio prompt or a text prompt as condition input, enabling flexible end-to-end extraction. The core of our proposed model lies in a tri-axial Transformer that jointly models temporal, frequency, and spatial channel dependencies. The model uses contrastive
language–audio pretraining (CLAP) embeddings to enable unified audio–text conditioning via feature-wise linear modulation (FiLM). To eliminate costly annotations and improve generalization, we propose a label-free data pipeline that dynamically generates spatial mixtures and corresponding targets for training. The result of our experiment with high separation quality demonstrates the efficacy of multimodal conditioning and tri-axial modeling. This work establishes a new paradigm for high-fidelity spatial audio separation in immersive applications.
\end{abstract}
\vspace{-1mm}
\section{Introduction}
\label{sec:intro}
Audio source separation~\cite{SS, MSS} is a fundamental problem in audio signal processing, aiming to recover individual sound events from complex mixtures. The task has recently gained renewed importance due to the growing demand for spatialized audio processing~\cite{rafaely2022spatial} in applications such as immersive media, augmented and virtual reality (AR/VR), hearing aids, and human–robot interaction. In these scenarios, separation models must not only achieve accurate source separation but also make effective use of spatial cues to distinguish direct sound from reverberation. This increasing demand calls for models that can operate reliably in reverberant, cluttered, and dynamic real-world acoustic environments.

Recent advances in deep learning have significantly improved separation performance in both monaural and stereophonic conditions. However, most existing approaches mainly emphasize time-domain modeling or time-frequency representations~\cite{convtasnet, 9413901, 10214650}, while insufficiently exploiting spatial cues that are essential to human auditory perception. Moreover, many separation systems are trained in a class-specific manner~\cite{seetharaman2019class, tzinis2020improving, pons2024gass} (e.g., focusing on speech or a definite set of sound events), which restricts their generalizability and hinders their applicability to diverse real-world scenarios.
Meanwhile, although some recent studies~\cite{LASS, audiosep, ma2024clapsep, 10887543} have explored target sound separation with multimodal cues, these efforts remain confined to single-channel audio and fail to exploit spatial information.
Conventional spatial filtering or beamforming methods~\cite{zhang2021adl} further struggle when spatial reverberation is strong. Consequently, designing a framework that can jointly capture temporal and spatial dependencies while also supporting end-to-end, query-based separation remains an open challenge.

In this paper, we propose the \textbf{B}and-split \textbf{S}patial \textbf{A}udio \textbf{S}eparation \textbf{T}ransformer (BSAST), a novel framework for query-based spatial audio extraction in reverberant and acoustically complex environments. Our model operates on first-order ambisonics (FOA) inputs, explicitly integrating spatial-channel cues alongside time-frequency representations to capture multi-dimensional dependencies. To enhance spectral modeling, we employ a band-split strategy~\cite{10447771}, dividing the input spectrum into multiple non-overlapping frequency bands. A tri-axial rotary positional encoding (RoPE) transformer~\cite{vaswani2017attention, rouard2023hybrid, su2024roformer} then applies attention sequentially along the time, frequency, and spatial–channel dimensions, enabling the model to capture complex interactions across all three axes. BSAST is designed to extract target sound events from reverberation and background interference, achieving robust and high-fidelity separation performance under realistic acoustic conditions.

Furthermore, our framework supports flexible, open-domain source extraction by accepting either audio exemplars or text descriptions as queries. We achieve this versatility through contrastive language–audio pretraining (CLAP) embeddings~\cite{10095969}, which inject semantic guidance directly into the extraction process. To scale training without reliance on strongly labeled data, we introduce a simple yet effective label-free data generation approach, where controlled noise is injected into the CLAP embeddings of target events to generate diverse query conditions on the fly. This strategy removes the need for paired audio–text annotations, substantially increases training diversity, and lowers the barrier to developing separation models in underexplored spatial audio scenarios.

Our main contributions are summarized as follows:
\begin{itemize}
    \vspace{-1mm}
    \item We propose BSAST, a spatial audio extraction framework that jointly models temporal, frequencial, and spatial-channel cues for robust extraction under reverberant conditions.
    \vspace{-1mm}
    \item We introduce a CLAP-based query conditioning mechanism, enabling both audio-based and text-based target source extraction beyond fixed-class settings.
    \vspace{-1mm}
    \item We design a label-free data pipeline that dynamically generates spatial mixtures and corresponding targets, improving training scalability.
\end{itemize}

\begin{figure*}[t]
    \centering
    \vspace{-2mm}
    \includegraphics[width=0.6\textwidth]{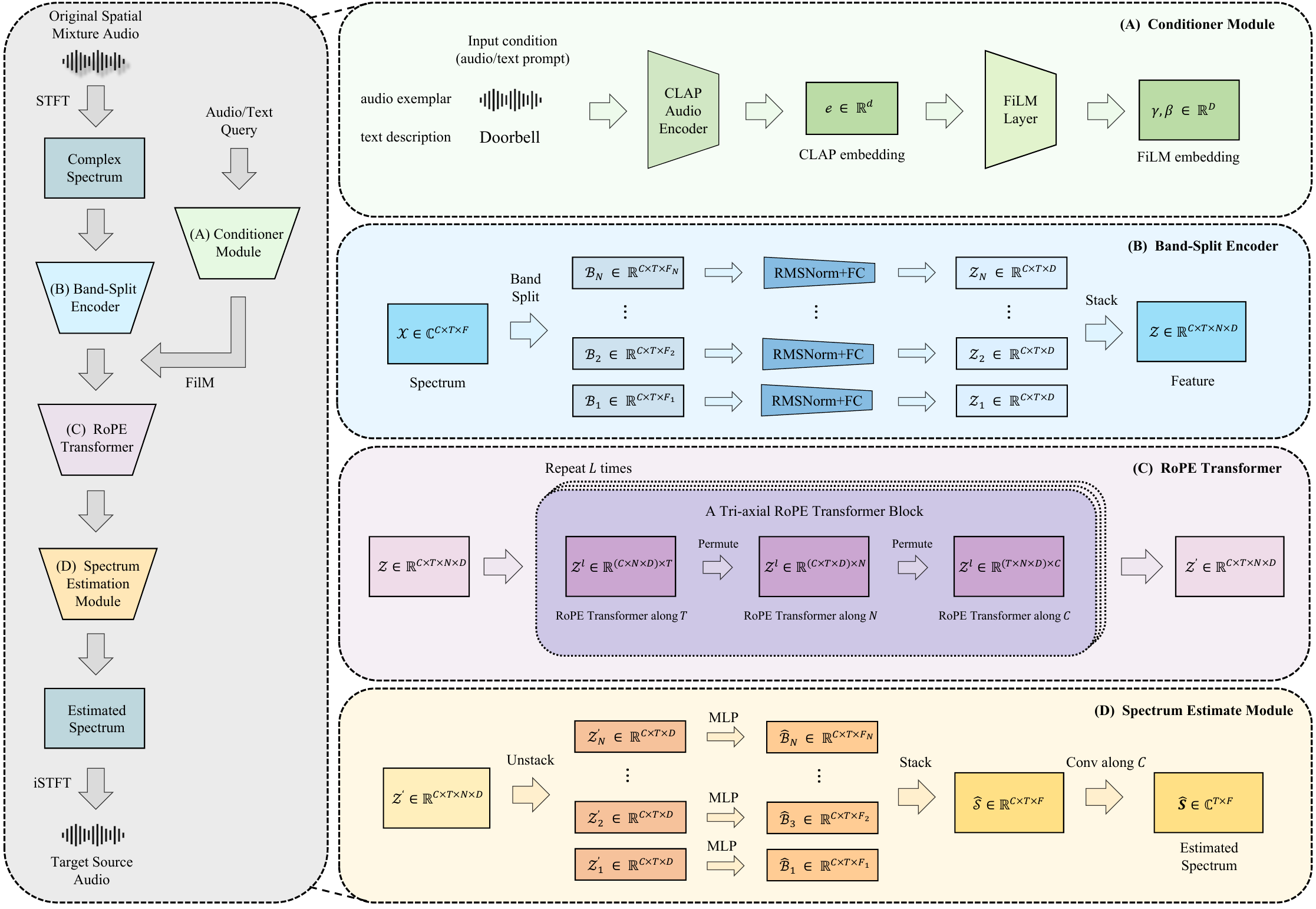}
    \vspace{-2mm}
    \caption{BSAST pipeline for query-based spatial audio extraction.}
    \vspace{-4mm}
    \label{fig:wide_figure}
\end{figure*}

\section{Problem Formulation}
\label{sec:formulation}
We address query-based spatial audio source extraction, where the goal is to recover a dry target signal from a multichannel mixture. The input mixture, given in first-order ambisonics (FOA) format, contains multiple simultaneously active sound events convolved with room impulse responses (RIRs) and overlaid with background noise and interference. The key challenge lies in disentangling overlapping sources in time and space, suppressing reverberation, and enabling flexible extraction of any target source specified by a query.

Formally, let 
$\mX \in \mathbb{R}^{C \times L}$
denote a multichannel FOA mixture, where $C$ and $L$ are the numbers of channels and audio samples, respectively. The mixture is generated as:
\begin{equation}
\mX = \sum_{i=1}^{M} \vs_{i} * \mH_{i} + \mN,
\end{equation}
where $M$ is the number of sources, $\vs_{i}$ are dry source signals, $\mH_{i}$ are corresponding multi-channel RIR, $*$ denotes convolution performed per channel after broadcasting $\vs_i$ to $C$ channels, and $\mN$ represents non-directional background noise and interference.

Given a query $\vq$ (either an audio exemplar or a text description), the goal is to estimate the corresponding dry target signal $\hat{\vs}_{q} \in \mathbb{R}^{T}$ from the mixture:
\begin{equation}
\hat{\vs}_{q} = f_{\theta}(\mX, \vq),
\end{equation}
where $f_{\theta}$ is a trainable model. This query-based formulation enables flexible source selection without prior knowledge of the mixture content. In this work, $f_{\theta}$ specifically refers to the proposed BSAST framework.

\section{Method}
\label{sec:method}

\subsection{System overview}
\label{ssec:overview}
Fig. \ref{fig:wide_figure} presents the overall system architecture.
Specifically, given a multichannel FOA mixture $\mX \in \mathbb{R}^{C \times T}$, the system first transforms the mixture into the time–frequency (T-F) representation $\mathcal{X} \in \mathbb{C}^{C \times T \times F}$ by applying the short-time Fourier transform (STFT) independently for each channel, where $T$ and $F$ are the number of frames and frequency bins, respectively. Subsequently, a band-split encoder module divides the spectrum into multiple subbands and extracts latent embeddings $\mathcal{Z}_n \in \mathbb{R}^{C \times T \times D}$ for each subband $n \in \{1, \dots, N\}$, where $N$ and $D$ are the numbers of subbands and latent features, respectively. We stack all $\mathcal{Z}_n$ along the subband axis to obtain a stack feature map $\mathcal{Z} \in \mathbb{R}^{C \times T \times N \times D}$. Meanwhile, the query condition is encoded using a pretrained CLAP encoder and injected via FiLM, producing query-adaptive feature maps. Then, the tri-axial RoPE Transformer models dependencies along the time, frequency, and channel dimensions. The output is denoted as $\mathcal{Z'} \in \mathbb{R}^{C \times T \times N \times D}$. Following tri-axial modeling, the spectrum estimation module predicts the target spectrum $\mathcal{\hat{S}}_q \in \mathbb{R}^{C \times T \times F}$ directly for each channel, and the channel merge module combines multi-channel outputs into a single-channel spectrum $\hat{\mS}_q \in \mathbb{C}^{T \times F}$. The final target waveform is obtained via iSTFT. During training, a label-free data construction pipeline generates diverse mixtures on-the-fly using dry sources, RIRs, and background noises, with pseudo-query embeddings derived from target audio CLAP embeddings.

\subsection{BSAST architecture}
\label{ssec:subhead}

Our BSAST backbone mainly contains a band-split encoder, a FiLM condition module, a tri-axial RoPE Transformer, and a spectrum estimate module.

\subsubsection{Band-split encoder}
\label{sssec:subsubhead}

Previous studies~\cite{10447771, su2024roformer} have demonstrated that splitting the input spectrum into multiple subbands can improve the performance while reducing computational complexity. The intuition is that the mixture signal spans the entire frequency range, where the low- and high-frequency components often exhibit distinct statistical and perceptual properties. By decomposing the spectrum into subbands, the model can better capture band-specific structures and mitigate interference across distant frequency regions. Building on this principle, we adopt a band-split module tailored for spatial audio. Specifically, we restructure the frequency division to better align with the multi-dimensional processing in the subsequent module. After STFT, we first map the complex spectrogram $\mathcal{X} \in \mathbb{C}^{C \times T \times F}$ to the real domain by separating and concatenating the real and imaginary parts along the frequency dimension.  Then the spectrogram is partitioned into $N$ uneven, non-overlapping subbands along the frequency axis to get $\mathcal{B}_n \in \mathbb{R}^{C \times T \times F_n}$, where $F_n$ denotes the number of frequency bins in band $n$. For each subband, we employ a lightweight feature extractor composed of RMS normalization~\cite{zhang2019root} and a linear projection, producing embeddings $\mathcal{Z}_n \in \mathbb{R}^{C \times T \times D}$, where $D$  is the hidden dimension. Finally, all subband embeddings are concatenated along the band axis, yielding the stacked representation $\mathcal{Z} \in \mathbb{R}^{C \times T \times N \times D}$, which is used as the input to the subsequent tri-axial RoFormer blocks.

\subsubsection{CLAP conditioning via FiLM}
\label{sssec:conditioning}

To enable query-based extraction, we employ semantic conditioning derived from CLAP embeddings~\cite{10095969}. CLAP provides a joint representation space for both audio and text queries, which allows our model to accept either an audio exemplar or a text description as conditioning input. A key design choice lies in how to inject these embeddings into the extraction backbone.

We adopted FiLM~\cite{perez2018film} to inject conditioning, as it provides a lightweight yet effective mechanism by modulating feature distributions without altering the backbone design. Concretely, given a CLAP embedding $\ve \in \mathbb{R}^{d}$, we first map it through a two-layer fully connected network with a ReLU activation, yielding a vector of size $2D$. This vector is then split into two parts, $\bm\gamma, \bm\beta \in \mathbb{R}^{D}$, where $D$ is the hidden dimension of the extraction backbone. For an intermediate feature map $\mathcal{Z} \in \mathbb{R}^{C \times T \times N \times D}$ from the band-split encoder, the modulation parameters are broadcast across non-feature dimensions and applied as
$
\ FiLM(\mathcal{Z}, \bm\gamma, \bm\beta) = \bm\gamma \odot \mathcal{Z} + \bm\beta,
$
where $\bm\gamma$ scales and $\bm\beta$ shifts the latent features, and $\odot$ denotes element-wise multiplication. In our implementation, the FiLM layer is inserted immediately after the band-split encoder and before each RoFormer block.

\subsubsection{Tri-axial RoPE Transformer blocks}
\label{sssec:transformer}
We adopt a tri-axial RoPE Transformer as the main extraction module. Each block sequentially applies axial attention along the time, frequency, and channel axes, enabling the model to explicitly model interactions and efficiently exchange information within and across these dimensions.  Additionally, we apply RoPE~\cite{su2024roformer} to encode relative positional dependencies along each axis.

Formally, given $\mathcal{Z} \in \mathbb{R}^{C \times T \times N \times D}$, which is the query-adaptive feature map after FiLM modulation, each Transformer block first normalizes the input using RMSNorm~\cite{zhang2019root} and encodes positional information via RoPE. Multi-head attention is then performed sequentially along each axis in the following order: time, frequency, and channel. When computing attention along a specific axis, the other dimensions are stacked together to form the input sequence. In a single Transformer block, a multi-head attention module is followed by a standard feedforward module, and both of them include residual connections. Stacking $L$ such blocks produces a final latent representation for downstream source extraction. This tri-axial RoPE Transformer design effectively integrates temporal dynamics, frequency bands structure, and spatial-channel dependencies in a unified framework.

\subsubsection{Spectrum estimation module}
\label{sssec:subsubhead}
After the tri-axial RoPE Transformer blocks, the model produces latent features that capture temporal, spectral, and spatial-channel dependencies. These features are then passed to the spectrum estimate module, which directly predicts the magnitude spectrum of the target source for each subband, rather than estimating a time-frequency mask. Such complex spectral mapping has been shown to improve convergence stability and reconstruction fidelity, especially in highly overlapped or reverberant mixtures~\cite{10214650}.

Specifically, given the output features from the tri-axial RoPE Transformer blocks $\mathcal{Z'} \in \mathbb{R}^{C \times T \times N \times D}$, we first unbind along the subband axis, yielding $\mathcal{Z'}_n \in \mathbb{R}^{C \times T \times D}$ for each band $n \in \{1, \dots, N\}$. Each band is processed by an MLP with gated linear units (GLU) to generate the estimated spectrum $\mathcal{\hat{B}}_{n} \in \mathbb{R}^{C \times T \times F_{n}}$. The outputs from all subbands are then concatenated along the frequency axis to obtain the estimated full-band spectrum $ \mathcal{\hat{S}} \in \mathbb{R}^{C \times T \times F}$. To aggregate information across channels, the concatenated spectrum is further passed through a channel merge Module, implemented as a convolution network with reduction to a single channel. This module effectively integrates multi-channel information while preserving the spatially-resolved spectral content, producing the final spectrum estimate $\hat{\mS} \in \mathbb{C}^{T \times F}$.

\subsection{Label-free training data construction}
\label{ssec:subhead}

To enable large-scale training without costly human annotations, we adopt a label-free data construction strategy inspired by recent advances in audio-text joint embedding models~\cite{10887543}. Specifically, we leverage a pretrained audio–text representation model to obtain audio embeddings from unlabeled source audio. These embeddings are then perturbed with controlled noise to generate pseudo-text query embeddings, where the injected perturbations simulate natural modality discrepancies between audio and language representations. This strategy yields training samples of query-target pairs without requiring explicit parallel labels. During inference, the model benefits from this design by supporting flexible query modalities. Compared with conventional supervised pipelines, this approach scales seamlessly to large unlabeled corpora while preserving cross-modal alignment.

\section{EXPERIMENTS}
\label{sec:pagestyle}

\subsection{Datasets}
\label{ssec:subhead}

We adopt the official dataset released for the detection and classification of
acoustic scenes and events (DCASE) 2025 Task 4~\cite{yasuda2025description}. The dataset consists of anechoic dry sources covering a wide range of sound event classes (Anechoic Sound Event 1K, FSD50K, and EARS dataset), room impulse responses (RIRs) recorded in FOA format, and nondirectional background noise and interference events (FOA-MEIR, FSD50K, ESC-50, DISCO). All audio is standardized to 32 kHz and 16-bit.

To synthesize spatialized mixtures, we employ the official Spatial Scaper~\cite{roman2024spatial}. For each mixture, dry source events are randomly selected and convolved with RIRs, then mixed with background noise and interference events. The resulting mixtures are 10 seconds in duration with up to three events overlapping simultaneously. For training, mixtures are dynamically generated by randomly sampling sources, RIRs, and noise from the dataset. This design substantially increases the training diversity. For evaluation, the test set contains 3,000 pre-mixed mixtures synthesized from previously unseen dry sources, RIRs, and noise.

\subsection{Training objectives}
\label{ssec:subhead}

Our training objective combines scale-invariant signal-to-distortion ratio (SI-SDR) loss with an $L_{1}$ reconstruction loss to balance perceptual quality and waveform fidelity.

For a predicted waveform $\hat{s}$ and target waveform $s$, the SI-SDR ~\cite{8683855} is defined as:
\begin{equation}
\text{SI-SDR}(\hat{s}, s) = 10 \log_{10} \frac{\|\alpha s\|^{2}}{\|\hat{s} - \alpha s\|^{2}}, \quad \text{where} \quad \alpha = \frac{\langle \hat{s}, s \rangle}{\|s\|^{2}}.
\end{equation}
Here $\alpha s$ denotes the target signal projected onto the direction of $\hat{s}$, removing scale ambiguity and making the metric invariant to gain. The SI-SDR loss is defined as the negative of the computed SI-SDR value. Additionally, to promote waveform fidelity in the reconstructed audio waveforms, an $L_{1}$ loss is incorporated as:
\begin{equation}
\mathcal{L}_{\text{SI-SDR}} = -\text{SI-SDR}(\hat{s}, s). \quad \mathcal{L}_1 = \|\hat{s} - s\|_1.
\end{equation}
Finally, the overall multi-objective loss is defined as a weighted sum:
\begin{equation}
\mathcal{L} = \mathcal{L}_{\text{SI-SDR}} + \lambda \mathcal{L}_{1},
\end{equation}
where $\lambda$ is a balancing coefficient. In our experiments, we set $\lambda = 100$ based on empirical tuning.

\subsection{Implementation details}
\label{ssec:subhead}

We perform STFT on each FOA channel with a Hanning window length of 2048 and a hop length of 1024. The frequency axis is partitioned into 25 non-overlapping subbands according to a pre-defined band-split scheme: 11 low-frequency bands of 6 bins each, followed by 6 mid-frequency bands of 32 bins each, 4 high-frequency bands of 64 bins each, and 3 ultra-high-frequency bands of 128, 128, 128, and 127 bins respectively, resulting in 25 bands in total.
The feature dimension is set to 128, and the backbone consists of 8 RoPE Transformer blocks. Each block employs 4 attention heads with a head dimension of 64. The spectrum estimator is implemented with a multi-layer perceptron (MLP) of depth 2 and an expansion factor of 4.
Optimization is performed using the AdamW~\cite{loshchilov2017decoupled} optimizer with an initial learning rate (LR) $3 \times 10^{-4}$ and a weight decay of $1 \times 10^{-2}$. Mixed-precision training is adapted to improve efficiency. The model is trained for a maximum of 300 epochs on an NVIDIA H100-80GB GPU with a batch size of 4. Each epoch contains 2,000 mixture samples. We apply gradient accumulation with an accumulation step of 2.

\subsection{Results}

We evaluate BSAST on the official test split using SI-SDR and SDR as evaluation metrics. The experiments are designed to assess the capability of our proposed method in multi-modal query-based spatial audio extraction. Results are reported under two types of query conditioning: a noise-added audio query and a clean text query.

\begin{table}[t]
\centering
\begin{tabular}{|c|c|c|c|c|}
\hline
\multirow{2}{*}{Channel} & \multicolumn{2}{c|}{Audio Condition} & \multicolumn{2}{c|}{Text Condition} \\
\cline{2-5}
       & SI-SDR & SDR & SI-SDR & SDR \\
\hline
wxyz (full FOA)  & \textbf{7.296} & \textbf{8.595} & \textbf{4.098} & \textbf{5.664} \\
\hline
w (omni channel only)  & 5.833 & 6.785 & \textbf{4.101} & 4.557 \\
\hline
\end{tabular}
\caption{Median SI-SDR and SDR performance (dB) on different channel configurations.}
\label{tab:chans}
\vspace{-4mm}
\end{table}

\begin{table}[ht]
\centering
\begin{tabular}{|c|c|c|c|c|}
\hline
\multirow{2}{*}{Blocks} & \multicolumn{2}{c|}{Audio Condition} & \multicolumn{2}{c|}{Text Condition} \\
\cline{2-5}
       & SI-SDR & SDR & SI-SDR & SDR \\
\hline
4      & 4.791 & 6.273 & 2.435 & 3.052 \\
\hline
6      & 6.426 & 7.752 & 3.871 & 4.459 \\
\hline
8      & \textbf{7.296} & \textbf{8.595} & \textbf{4.098} & \textbf{5.664} \\
\hline
\end{tabular}
\caption{Median SI-SDR and SDR performance (dB) across different Transformer block configurations.}
\label{tab:depth_si_sdr}
\end{table}

\begin{figure}[t] 
    \centering
    \vspace{-4mm}
    \includegraphics[width=0.8\linewidth]{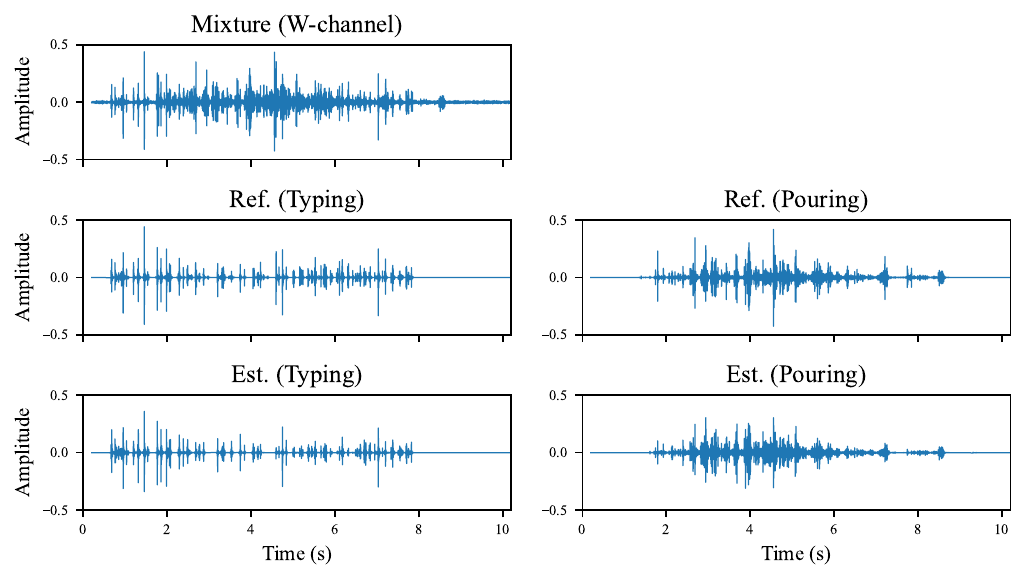}
    \vspace{-4mm}
    \caption{Example of an estimated target using text query.} 
    \vspace{-6mm}
    \label{fig:showcase}
\end{figure}

In Table~\ref{tab:chans}, we first evaluate the model’s ability to exploit spatial cues for audio source extraction. We trained two versions of BSAST: a full version that takes the four FOA channels (wxyz) as input, thereby leveraging the spatial information in multi-channel audio, and an ablated version that uses only the single omnidirectional channel (w), corresponding to the absence of spatial cues. Experimental results demonstrate that the model achieves better performance when complete spatial information is available, thereby validating the effectiveness of the proposed time–frequency–spatial modeling strategy in capturing temporal, frequential, and spatial dependencies for separating complex spatial audio mixtures under reverberant conditions.

In Table~\ref{tab:depth_si_sdr}, we then examine the influence of model depth. Increasing the number of Transformer blocks leads to steady performance gains, with the best results obtained with 8 blocks. This trend highlights the scalability of BSAST, showing that deeper architectures can better capture the intricate structures of complex spatial audio mixtures. Moreover, the consistent improvements suggest that the model design can effectively leverage additional capacity without signs of saturation, pointing to strong robustness and potential for further scaling.

Notably, although BSAST is trained exclusively with audio queries using the label-free data construction strategy, it also achieves competitive results when tested with clean text queries. This demonstrates the versatility of the CLAP-based FiLM conditioning, which enables the framework to support both audio- and text-driven extraction beyond fixed-class settings. Fig.~\ref{fig:showcase} illustrates a representative test example, showing that BSAST can recover individual sources with high fidelity and clarity.

Together, these results underscore the three core innovations of BSAST—time–frequency–spatial dependency modeling, CLAP-based multi-modal query conditioning, and a label-free training pipeline—while also establishing a robust and extensible foundation for high-fidelity spatial source extraction. Building on this foundation, further scaling of both model capacity and training data is expected to yield even stronger performance across diverse and challenging acoustic scenes.

\section{CONCLUSION}
\label{sec:typestyle}
We introduced a flexible framework for query-based spatial audio extraction that enables direct dry source recovery from multichannel mixtures using either audio or text queries. By unifying spatial, temporal, and spectral reasoning with modality-agnostic query conditioning, our approach simplifies source extraction without relying on predefined classes or labeled datasets. The proposed label-free data construction further enhances practicality, making it well-suited for real-world acoustic scenarios. This work demonstrates the promise of query-based extraction for advancing interactive and immersive audio technologies.

\bibliographystyle{IEEEbib}
\bibliography{strings,refs}

\end{document}

%% file: define.tex
\usepackage{amsmath,amsfonts,bm}

\def\eqref#1{equation~\ref{#1}}

\def\1{\bm{1}}

\def\ve{{\bm{e}}}

\def\vq{{\bm{q}}}

\def\vs{{\bm{s}}}

\def\mH{{\bm{H}}}

\def\mN{{\bm{N}}}

\def\mS{{\bm{S}}}

\def\mX{{\bm{X}}}

\DeclareMathAlphabet{\mathsfit}{\encodingdefault}{\sfdefault}{m}{sl}
\SetMathAlphabet{\mathsfit}{bold}{\encodingdefault}{\sfdefault}{bx}{n}

